\documentclass[letterpaper,10pt, conference]{ieeeconf}
\IEEEoverridecommandlockouts
\usepackage{cite}
\usepackage{amsthm}
\usepackage{amsmath}
\usepackage{algorithm}
\usepackage[varvw]{newtxmath}
\usepackage{epsfig} 
\usepackage{algcompatible}
\usepackage{subfig}
\usepackage{soul}
\usepackage{xcolor}

\def\BibTeX{{\rm B\kern-.05em{\sc i\kern-.025em b}\kern-.08em
T\kern-.1667em\lower.7ex\hbox{E}\kern-.125emX}}

\usepackage[top=0.75in,bottom=0.8in,right=0.75in,left=0.75in]{geometry}

\newcommand{\norm}[2]{\left \lVert #1 \right \rVert_{#2}}

\algnewcommand\algorithmicreturn{\textbf{return}}
\algnewcommand\RETURN{\State \algorithmicreturn}
\algrenewcommand\algorithmicrequire{\textbf{Input:}}
\algrenewcommand\algorithmicensure{\textbf{Output:}}

\newtheorem{theorem}{Theorem}

\newtheorem{corollary}{Corollary}

\newtheorem{definition}{Definition}
\newtheorem{remark}{Remark}

\begin{document}

\title{Collision Avoidance for Non-Cooperative Multi-Swarm Coverage Control with Bounded Disturbance Measurements}
\author{
Karolina Schmidt and Luis Rodrigues
\\
\small Department of Electrical and Computer Engineering \\
\small Concordia University, Montr\'eal, QC, Canada\\
\small Email: {\tt karolina.schmidt@mail.concordia.ca, luis.rodrigues@concordia.ca}
}

\maketitle

\begin{abstract}
This paper proposes a new algorithm for collision-free coverage control of multiple non-cooperating swarms in the presence of bounded disturbances. A new methodology is introduced that accounts for uncertainties in disturbance measurements. The proposed methodology is used to develop an algorithm that ensures collision-free motion in multi-swarm coverage control, specifically for cases where disturbances are present and their measurements are subject to bounded uncertainty. The theoretical results are validated through simulations of multiple swarms that independently aim to cover a given region in an environment with disturbances.
\end{abstract}

\noticetopractitioners 
    The motivation of this paper arises from the increased use of multi-agent systems over the past years, which has led to a need to operate multiple independent swarms within shared spaces. As outlined in the introduction section, examples include wildfire management, planetary exploration, and mobile coverage provided by drones. Important considerations to ensure safe motion are the avoidance of collisions between agents, as well as the compensation for disturbances, such as winds or currents, and uncertainty in their measurements. This paper proposes a new methodology for multi-agent collision avoidance with disturbance measurement uncertainty and mathematically guarantees collision-free motion. While the methodology can be used for general types of multi-agent systems, we propose a new algorithm that incorporates the collision avoidance methodology into coverage control with multiple non-cooperating swarms. Three-dimensional simulations show that, despite collision avoidance maneuvers, maximum coverage is achieved by all swarms.
\endnoticetopractitioners

\section{Introduction}
Due to the various tasks that swarms of agents are intended to execute, there is a need for multiple non-coorperating swarms to independently provide maximum coverage to a common region. In these scenarios, agents need to avoid collisions while compensating for unintended drifts caused by disturbances. Examples include wind drift for Unmanned Aerial Vehicles (UAVs) or currents in scenarios involving Autonomous Underwater Vehicles (AUVs). Applications for the provision of coverage with multiple non-cooperating swarms are search and rescue while simultaneously performing fire monitoring, planetary exploration where swarms belonging to different countries independently aim to cover as much terrain as possible, as well as mobile coverage in remote areas provided by drones from different companies. Motivated by these applications we propose an uncertainty-aware collision avoidance approach for multi-agent systems and apply it to coverage control with multiple non-cooperating swarms. The focus of the paper is on bounded uncertainties in the measurement of disturbances. Preliminary work by the same authors for the disturbance-free two-dimensional problem can be found in reference \cite{SchmidtRodrigues2024}.

Coverage control using centroidal Voronoi configurations was first investigated by Cortés et al. in references \cite{CortesMartinez1,CortesMartinez2}. Using Lloyd's iterative algorithm \cite{LloydsAlgorithm} and gradient descent methods, a swarm of agents is coordinated to provide maximum coverage for a region. Extensions and further advancements are proposed in reference \cite{MoarrefRodrigues2014} where energy conservation and efficiency are considered by formulating an optimal control problem. Moreover, reference \cite{DipernaRodrigues2017} introduces an analytic expression for the rate of change of the mass and the location of the center of mass of a Voronoi cell, while the authors in \cite{NguyenRodriguesManiuOlaru2016} address the problem introduced in \cite{MoarrefRodrigues2014} for discrete-time systems. More recent work addresses obstacle avoidance in coverage control using control barrier functions \cite{Chaitra2023,Ma2025}, leader-follower approaches \cite{BaiWang2022}, as well as ant colony algorithms \cite{YangLiang2024}. Position uncertainty in collision-free coverage control is addressed in \cite{SongFan2020} and \cite{ChenDames2020}, while \cite{BaiWang2023} considers system uncertainty. Various papers \cite{HaghighiCheah2012,SharifiPhDThesis2014,AtinEtAl2020,WangEtAl2021,ABBASI2017342} consider more than one swarm of agents in different scenarios. These include formation flying \cite{HaghighiCheah2012}, collaborative coverage control of UAVs and Unmanned Ground Vehicles \cite{SharifiPhDThesis2014}, time-efficient coverage control with cooperating swarms using a leader-follower approach \cite{AtinEtAl2020}, and the division of a swarm into sub-groups to increase coverage speed and reduce traveling distances \cite{WangEtAl2021} or to deploy heterogeneous
agents \cite{ABBASI2017342}. A review of coverage control algorithms is given in \cite{coveragecontrolreview2025}.

A widely studied approach to avoid moving obstacles for robot motion is the Velocity Obstacles (VO) method presented in \cite{VelocityObstacles1998}. Extensions and improvements include \cite{ReciprocalVelocityObstacles2008} and \cite{ClearPathFVO2009}, where collision-free motion for multi-agent systems is addressed. Reference \cite{ORCA2011} introduces a geometric approach called Optimal Reciprocal Collision Avoidance (ORCA) allowing the agents to solve a low-dimensional linear program to compute collision-free velocities. A detailed overview of VO-based collision avoidance methods is given in \cite{VelocityObstaclesReview2024}. Uncertainty in the position and velocity of obstacles in collision avoidance methods based on Velocity Obstacles is addressed in \cite{EllingsonPitts2020,GyenesSzadeczky2020,RodriguezErick2011,dergachev2025decentralizeduncertaintyawaremultiagentcollision}. Reference \cite{zhang2022velocity} presents risk-bounded motion planning using Velocity Obstacles with disturbances, Gaussian sensor noise, and model uncertainty. Nevertheless, none of the above references considers the effects of uncertainty in the measurements of disturbances. Independent of Velocity Obstacles and its extensions, reference \cite{BodduSuman2024} analyzes parameters in obstacle avoidance under the presence of wind.

To the best of the authors' knowledge, reference \cite{SchmidtRodrigues2024} is the only work that considers the scenario of multiple non-cooperating swarms that independently aim to cover a common region while avoiding collisions among agents, but it does not consider uncertainty. Moreover, the reference only considers planar motion of the agents, so that collision avoidance maneuvers are only performed in two dimensions. This can be a limiting factor when the agents are UAVs or AUVs. The contributions of this paper are the following:
\begin{enumerate}
    \item A methodology for collision-free motion in multi-agent scenarios accounting for uncertainty in the measurement of bounded disturbances is proposed. Formal guarantees for collision avoidance are derived.
    \item A new algorithm for collision-free coverage control with multiple non-cooperating swarms accounting for bounded disturbances and uncertainty in their measurements is proposed. Simulations are shown for three-dimensional motion, thus enabling UAV and AUV applications.
\end{enumerate} 
Notice that contribution 1 is not specific to non-cooperative multi-swarm coverage control as the method can be used for general types of multi-agent systems. It extends ORCA to the case of bounded disturbances with measurement uncertainty.

The paper is organized as follows. Section \ref{section:preliminaries} summarizes preliminaries on coverage control and ORCA. The proposed methodology to account for uncertainties in the measurements of disturbances is introduced and formulated in section \ref{section:proposedmethod}. It is then used to propose an algorithm for multi-swarm coverage control that is presented in section \ref{subsection:algorithm}. Three-dimensional simulations are shown in section \ref{section:simulations}, before conclusions are drawn in section \ref{section:conclusions}.

\section{Preliminaries} \label{section:preliminaries}

\subsection{Energy-Efficient Coverage Control} \label{subsection:preliminariescoveragecontrol}

This subsection follows references \cite{CortesMartinez1,CortesMartinez2,MoarrefRodrigues2014}. Let $Q$ be a convex region and let $n$ agents be located at distinct positions $X=\{x_1,...,x_n\}$, where $x_i \in Q, i \in \{1,...,n\}$. Then, using Voronoi tessellation, $Q$ can be divided into subsets called Voronoi cells $V_i$ so that each point $q \in Q$ is associated with the closest agent, i.e., $V_i(X)= \{ q \in Q \, | \norm{q-x_i}{2} \leq \norm{q-x_j}{2}\forall x_j \in X \}$. A distance function for a point $q \in Q$ is described as $f(x_{i(q)},q)={||x_{i(q)}-q||}$, where $x_{i(q)}$ is the position of the closest agent to $q$. High distance values correspond to a low level of coverage, while low values indicate high coverage. The importance of coverage at $q$ is described by a density function $\phi (q)$. Therefore, to maximize coverage the following function is minimized \cite{MoarrefRodrigues2014},
\begin{equation} \label{equation:lyapunovfunctionV_i}
    V(x) = \sum_{i=1}^{n} \int_{V_i} {||x_{i}-q||}^2 \phi(q) dq.
\end{equation}
Using the necessary condition for a local minimum yields
\begin{equation} \label{equation:necessarycondition_short}
    \frac{\partial V}{\partial x_i} = 2M_{V_i} (x_i-CM_{V_i})^T = 0,
\end{equation}
where $M_{V_i}$ is the mass and $CM_{V_i}$ is the center of mass of Voronoi cell $V_i$, written as
\begin{equation} \label{equation:mass}
    M_{V_i} = \int_{V_i} \phi (q) dq,
\end{equation}
\begin{equation} \label{equation:centerofmass}
    CM_{V_i} = \frac{\int_{V_i}q \phi (q)dq}{\int_{V_i} \phi (q)dq} = \frac{\int_{V_i}q \phi (q)dq}{M_{V_i}}.
\end{equation}
Equation (\ref{equation:necessarycondition_short}) is satisfied when $x_i=CM_{V_i}$. A configuration where each agent is located at the center of mass of its Voronoi cell is called a centroidal Voronoi configuration. Lloyd's iterative algorithm \cite{LloydsAlgorithm} can be used to move each agent towards the center of mass of its Voronoi cell.

The authors of \cite{MoarrefRodrigues2014} formulate the efficient coverage task as the following optimal control problem,
\begin{align} \label{equation:optimalcontrolproblemwithenergy}
    \begin{split}
        \underset{u_i,i\in\{1,...,n\}}{inf} & \int_{0}^{\infty} \sum_{i=1}^{n} \left( s_i ||\int_{V_i}(x_i-q)\phi(q)dq||^2+r_iu_i^Tu_i \right) d \tau\\
        s.t. \; & \dot{x}_i=u_i,
    \end{split}
\end{align}
where $u_i$ is the control input of agent $i$. The coefficients $s_i \geq 0$ and $r_i>0$ weigh the importance of coverage as measured by
\begin{equation} \label{equation:coveragecriterion}
    M_{V_i}||x_i-CM_{V_i}||=||\int_{V_i}(x_i-q)\phi(q)dq||
\end{equation}
and input energy for each agent, respectively. The authors in \cite{MoarrefRodrigues2014} prove that 
a solution of (\ref{equation:optimalcontrolproblemwithenergy}) is
\begin{equation} \label{equation:controlinputwithenergy}
    u_i = - \sqrt{s_i/r_i} \int_{V_i} (x_i-q) \phi (q) dq.
\end{equation}
A first order dynamical model for the closed-loop dynamics of each agent is thus 
\begin{equation} \label{equation:dynamicsystem}
    \begin{split}
        & \dot{x}_i = u_i, \\
        & u_i = k_i(CM_{V_i}-x_i),
    \end{split}
\end{equation}
where $k_i=M_{V_i}\sqrt{s_i/r_i}$.

\subsection{Optimal Reciprocal Collision Avoidance (ORCA)} \label{subsection:originalorca}
This subsection follows reference \cite{ORCA2011}. Consider two spherical agents $A$ and $B$ with radii $r_A$ and $r_B$ and centered at positions $x_A$ and $x_B$, respectively. 

\begin{definition}
    A collision between two spherical agents $A$ and $B$ occurs, when
    \begin{equation} \label{equation:collisiondefinition}
        \norm{x_A - x_B}{} \leq r_A + r_B.
    \end{equation}
    Conversely, the agents do not collide when
    \begin{equation} \label{equation:nocollisiondefinition}
        \norm{x_A - x_B}{} > r_A + r_B.
    \end{equation}
\end{definition}

\begin{figure}[t]
    \centering
    \includegraphics[scale=0.46]{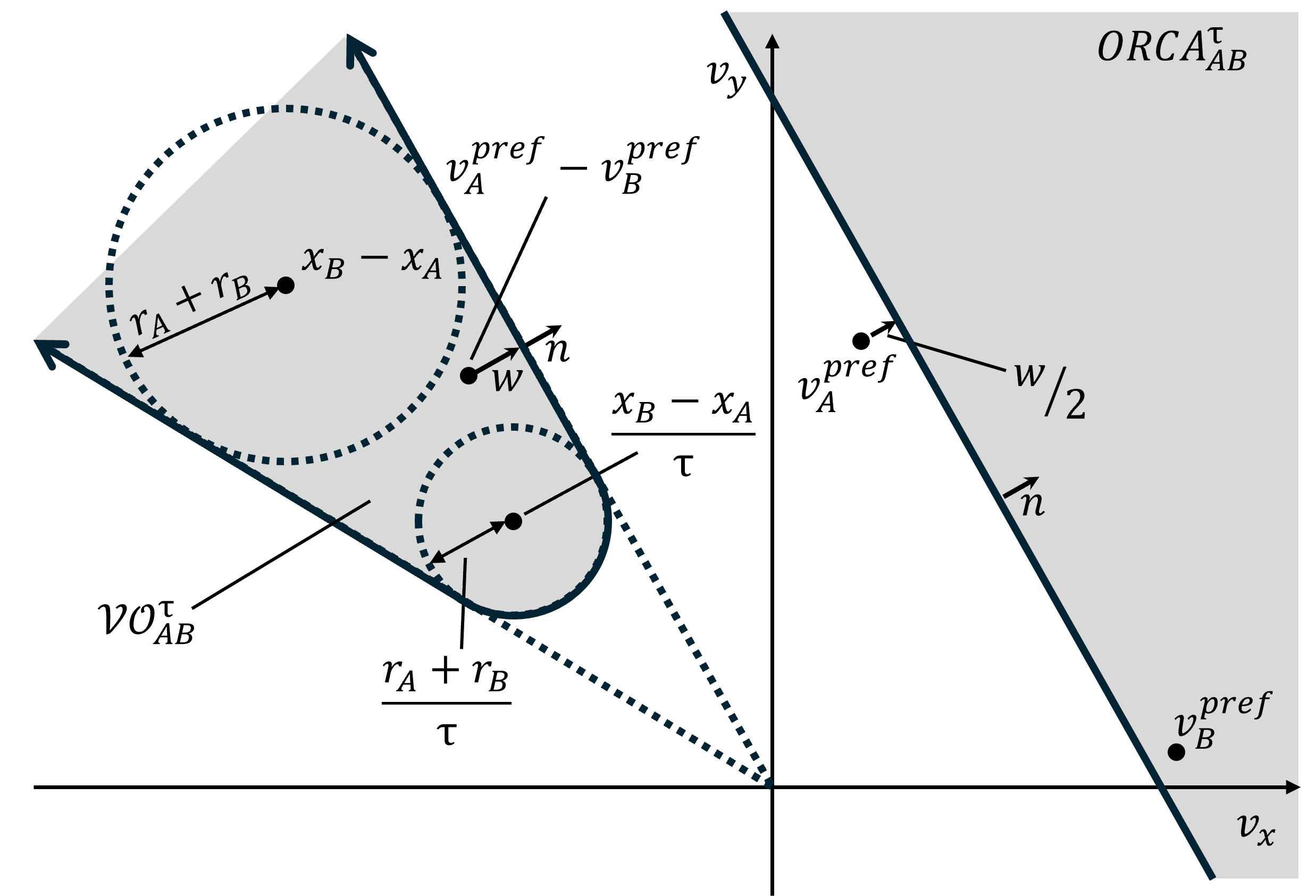}
    \caption{Geometry of $\mathcal{VO}_{AB}^{\tau}$, adapted from \cite{ORCA2011}}
    \label{fig:orca}
\end{figure}

Using ORCA, agent $A$ can compute a truncated collision cone in the velocity space denoted as $\mathcal{VO}_{AB}^{\tau}$ and called the Velocity Obstacle of $A$ with respect to $B$ for time horizon $\tau$. The Velocity Obstacle $\mathcal{VO}_{AB}^{\tau}$ is computed as \cite{ORCA2011}
\begin{equation} \label{equation:formalvelocityobstacle}
    \mathcal{VO}_{AB}^{\tau} = \left\{ v \, | \, \exists t \in (0, \tau ]:v \in S((x_B-x_A)/t,(r_A+r_B)/t)\right\},
\end{equation}
where $S$ describes a sphere,
\begin{equation} \label{equation:disc}
    S(x,r) = \left\{ q \left| \ \norm{q-x}{} \leq r \right. \right\}.
\end{equation}
The set $\mathcal{VO}_{AB}^{\tau}$ in two dimensions is shown in figure \ref{fig:orca}.

\begin{theorem} \label{theorem:collisioniffinVO}
    (\cite{thesisschmidt}, page 24) Consider two agents $A$ and $B$ traveling with velocities $v_A$ and $v_B$, respectively. Let both $v_A$ and $v_B$ be constant for a time horizon $\tau$. Then, $A$ and $B$ collide within the time horizon $\tau$ if and only if
    \begin{equation} \label{equation:relativevelinvo}
        v_A-v_B \in \mathcal{VO}_{AB}^{\tau}.
    \end{equation}
    Therefore, $A$ and $B$ do not collide if and only if
    \begin{equation} \label{equation:relativevelnotinvo}
        v_A-v_B \notin \mathcal{VO}_{AB}^{\tau}.
    \end{equation}
\end{theorem} 
\vspace{0.5\baselineskip}
Let $v_A^{pref}$ and $v_B^{pref}$ be the desired velocities that $A$ and $B$ aim to follow, respectively, to meet other potential path planning objectives. To avoid a collision with $B$, $A$ computes the shortest vector $w$ from 
\begin{equation} \label{equation:relativevelocitynew}
    v_{AB}^{pref}=v_{A}^{pref}-v_{B}^{pref}
\end{equation}
to the boundary of $\mathcal{VO}_{AB}^{\tau}$, denoted as $\partial \mathcal{VO}_{AB}^{\tau}$ \cite{ORCA2011},
\begin{equation} \label{equation:shortestvectortoboundary}
    w = \left( \underset{v \in \partial \mathcal{VO}_{AB}^{\tau}}{\text{argmin}} \norm{v-v_{AB}^{pref}}{}\right) - v_{AB}^{pref}.
\end{equation} 
Assuming that $A$ and $B$ share the responsibility to avoid each other, both agents compute a half-space containing velocities that they can adopt so that, if both $A$ and $B$ keep their velocities constant during a time horizon $\tau$, the two agents do not collide with each other within $\tau$ instants of time. The half-space of agent $A$ with respect to $B$ is described as \cite{ORCA2011}
\begin{equation} \label{equation:halfplane}
    ORCA_{AB}^{\tau} = \left\{ v \, \left| \, \left(v-\left(v_A^{pref}+ \frac{1}{2} w\right)\right) \cdot n > 0 \right. \right\},
\end{equation}
where $n$ is the outward normal vector of $\partial \mathcal{VO}_{AB}^{\tau}$ at $v_{AB}^{pref} + w$. The half-space of agent $B$ with respect to $A$ is
\begin{equation} \label{equation:orca_ba_halfplane}
    ORCA_{BA}^{\tau} = \left\{ v \, \left| \, \left(v-\left(v_B^{pref} - \frac{1}{2} w \right) \right) \cdot \left(-n\right) > 0 \right. \right\}.
\end{equation}
Notice that the boundary of $ORCA_{BA}^{\tau}$ is parallel to that of $ORCA_{AB}^{\tau}$, but the corresponding normal vector points in the opposite direction.
\begin{theorem} \label{theorem:nocollisionifvnewanywhereinhp}
    (\cite{thesisschmidt}, page 33) Consider an agent $A$ and an agent $B$. Let $A$ adopt any velocity $v_A^{new} \in ORCA_{AB}^{\tau}$ and let $B$ adopt any velocity $v_B^{new} \in ORCA_{BA}^{\tau}$. If $v_A^{new}$ and $v_B^{new}$ are constant for a time horizon $\tau$, then $A$ and $B$ do not collide before $\tau$, i.e.,
    \begin{equation} \label{theorem3:relative new velocities not in VO}
        v_A^{new}-v_B^{new} \notin \mathcal{VO}_{AB}^{\tau}.
    \end{equation}
\end{theorem}
\vspace{0.5\baselineskip}
In multi-agent scenarios with $m$ agents, the set of feasible velocities for $A$ is obtained by computing a half-space with respect to each other agent $B_1,...,B_{m-1}$ along with a sphere $S(0,v_A^{max})$, where $v_A^{max}$ is the maximum speed achievable by $A$. Taking the intersection of all half-spaces computed by $A$ and $S(0,v_A^{max})$ yields
\begin{equation} \label{equation:intersectionofhalfplanes}
    ORCA_{A}^{\tau} = S(0,v_A^{max}) \cap \bigcap_{i=1}^{m-1} ORCA_{AB_i}^{\tau}.
\end{equation}
\begin{corollary}
    (\cite{thesisschmidt}, page 35) Consider $m$ agents operating within the same area and following the ORCA approach to avoid collisions with each other. Then an agent $A$ choosing its new velocity $v_A^{new} \in ORCA_{A}^{\tau}$, where
    \begin{equation} \label{equation:intersectionofhalfplanesagain}
        ORCA_{A}^{\tau} = S(0,v_A^{max}) \cap \bigcap_{i=1}^{m-1} ORCA_{AB_i}^{\tau},
    \end{equation}
    will not collide with any other agent $B_i, i\in \left\{ 1,...,m-1 \right\}$, before a time horizon $\tau$, if $v_{B_i}^{new} \in ORCA_{B_i}^{\tau}$, assuming that $v_A^{new}$ and $v_{B_i}^{new}$ are constant for $\tau$, i.e.,
    \begin{equation} \label{corollary1:condition}
        v_A^{new}-v_{B_i}^{new} \notin \mathcal{VO}_{AB_i}^{\tau}, \forall i \in \left\{ 1,...,m-1 \right\}.
    \end{equation}
\end{corollary} 
\vspace{0.5\baselineskip}
Reference \cite{ORCA2011} specifies that because the velocity obstacle is a truncated cone, for a finite time horizon $\tau$, the origin of the velocity space always lies outside the velocity obstacle. $ORCA_{A}^{\tau}$ will therefore not be empty if the desired velocities of all agents are set to zero. A collision-free velocity $v_A^{new}$ is then computed as 
\begin{equation} \label{equation:ORCAnewvelocity}
    v_A^{new} = \underset{v \in ORCA_{A}^{\tau}}{\text{argmin}} ||v-v_A^{pref}||.
\end{equation}

\section{Proposed Method} \label{section:proposedmethod}
This section presents the main results of this paper and is divided into three subsections. Subsection \ref{subsection:disturbance} describes the disturbances and measurement uncertainty considered in this work. In subsection \ref{subsection:math}, a methodology that ensures collision-free motion when uncertainties are present is proposed and formalized. An algorithm for collision-free multi-swarm coverage control in the presence of disturbances is introduced in subsection \ref{subsection:algorithm}.

\subsection{Disturbance and Measurement Uncertainty} \label{subsection:disturbance}

Consider a first-order system where a velocity control input $u_i$ is applied to an agent $i$ operating in the presence of a disturbance. Denote the velocity of the disturbance at the location $x_i$ at time $t$ as $v_{d(i,t)}$. Then, the kinematics are
\begin{align} \label{equation:systemdynamicswithdisturbance}
    \begin{split}
        &\dot{x}_i = v_i,\\
        &v_i = u_i + v_{d(i,t)}.
    \end{split}
\end{align}
In this work, the measured disturbance velocity at the location $x_i$ at time $t$ is denoted as $v_{d(i,t)}^{meas}$. It is described by the true disturbance velocity plus an additive sensor bias $b_{d(i)}$ and time-dependent noise $e_{d(i,t)}$,
\begin{equation} \label{equation:measuredwind}
    v_{d(i,t)}^{meas} = v_{d(i,t)} + b_{d(i)} + e_{d(i,t)}.
\end{equation}
It is assumed that an estimate $\hat{b}_{d(i)}$ for $b_{d(i)}$ is available. We will describe the deviation of $\hat{b}_{d(i)}$ from $b_{d(i)}$ as
\begin{equation} \label{equation:deltabias}
    \delta b_{d(i)} = b_{d(i)}-\hat{b}_{d(i)}.
\end{equation}
To obtain an estimate $\hat{v}_{d(i,t)}$ for $v_{d(i,t)}$ based on the measured value and the estimated sensor bias, let
\begin{equation} \label{equation:subtractedbias}
    \hat{v}_{d(i,t)} = v_{d(i,t)}^{meas} - \hat{b}_{d(i)}.
\end{equation}
Inserting $v_{d(i,t)}^{meas}$ as defined in (\ref{equation:measuredwind}) into (\ref{equation:subtractedbias}) yields
\begin{equation} \label{equation:vhat2}
    \hat{v}_{d(i,t)} = v_{d(i,t)} + b_{d(i)} + e_{d(i,t)} - \hat{b}_{d(i)},
\end{equation}
which using equation (\ref{equation:deltabias}) leads to
\begin{equation} \label{equation:vhat3}
    \hat{v}_{d(i,t)} = v_{d(i,t)} + \delta b_{d(i)} + e_{d(i,t)}.
\end{equation}
Let
\begin{equation} \label{equation:xidefinition}
    \xi_{d(i,t)}=\delta b_{d(i)}+e_{d(i,t)}.
\end{equation}
It is assumed that
\begin{equation} \label{equation:errorsinellipses}
    \xi_{d(i,t)} \in \varepsilon_i
\end{equation}
is bounded by an ellipsoid, where
\begin{equation} \label{equation:ellipsedefinition}
    \varepsilon_i = \left\{z \left| z^TPz \leq 1 \right. \right\}
\end{equation}
for some given $P=P^T > 0$ and $z=[v_x\;v_y\;v_z]^T$. Equations (\ref{equation:vhat3}) and (\ref{equation:xidefinition}) yield
\begin{equation} \label{equation:vhat4rearr}
    \xi_{d(i,t)} = \hat{v}_{d(i,t)} - v_{d(i,t)}. 
\end{equation}
Given a desired velocity $v_i^{pref}$ that is to be adopted, we will now design a velocity control input $u_i$ that counteracts external disturbances based on their bounds. Assuming that sufficient control input to counteract these disturbances can always be achieved, consider
\begin{equation} \label{equation:desiredcontrolinputwithwindrewritten}
    u_i = v_i^{pref} - \hat{v}_{d(i,t)}.
\end{equation}
Inserting (\ref{equation:desiredcontrolinputwithwindrewritten}) into (\ref{equation:systemdynamicswithdisturbance}), we obtain
\begin{equation} \label{velocitydeviation}
    v_i = v_i^{pref} + v_{d(i,t)} - \hat{v}_{d(i,t)}.
\end{equation}
Using (\ref{equation:vhat4rearr}) and (\ref{velocitydeviation}) yields
\begin{equation} \label{truevelocityredone}
    v_i = v_i^{pref} - \xi_{d(i,t)}.
\end{equation}
Subsection \ref{subsection:math} addresses the effects of $\xi_{d(i,t)}$ by computing a new collision-free velocity $v_i^{new}$ based on $v_i^{pref}$ and $\varepsilon_i$.

\subsection{Collision Avoidance with Disturbance Measurements} \label{subsection:math}
In this subsection, a methodology based on ORCA is proposed to account for bounded measurement uncertainty and guarantee collision-free motion. We use the alphabet to identify each agent $i$, i.e., $i=A,B,...$, to be consistent with the notation in reference \cite{ORCA2011} as presented in section \ref{subsection:originalorca}.

Consider the Minkowski sum of $\varepsilon_A$ and $\varepsilon_B$ which is
\begin{equation} \label{minksumnottranslated}
    \varepsilon_{AB}^{sum} = \varepsilon_A + \varepsilon_B = \{ \eta_A+\eta_B \: | \: \eta_A \in \varepsilon_A, \eta_B \in \varepsilon_B \},
\end{equation}
and translate the resulting set $\varepsilon_{AB}^{sum}$ by $v_{AB}^{pref}$, which yields
\begin{equation} \label{equation:minkowskisumellipses}
    \overline{\varepsilon_{AB}^{sum}} = \varepsilon_{AB}^{sum} + v_{AB}^{pref}.
\end{equation}

\begin{theorem} \label{theorem:allvelocitiesineab}
    Assume that the constraint (\ref{equation:errorsinellipses}) is satisfied for both $\xi_{d(A)}$ and $\xi_{d(B)}$. If the agents $A$ and $B$ apply control inputs $u_A$ and $u_B$ as defined in equation (\ref{equation:desiredcontrolinputwithwindrewritten}) with $i$ replaced by $A$ and $B$, respectively, based on given desired velocities $v_A^{pref}$ and $v_B^{pref}$, then
    \begin{equation} \label{equation:vaminusvbinepsilonab}
        v_A - v_B \in \overline{\varepsilon_{AB}^{sum}},
    \end{equation}
    where $v_A$ and $v_B$ are given by equation (\ref{truevelocityredone}) with $i$ replaced by $A$ and $B$, respectively.
\end{theorem}

\proof
    From equation (\ref{truevelocityredone}) with $i$ replaced by $A$ and $B$, respectively, we can write
    \begin{align} \label{truevelocitiesaandb}
        \begin{split}
            &v_A = v_A^{pref} - \xi_{d(A)},\\
            &v_B = v_B^{pref} - \xi_{d(B)}.
        \end{split}
    \end{align}
    Subtracting $v_B$ from $v_A$ yields
    \begin{equation}
        v_A - v_B = v_A^{pref} - \xi_{d(A)} - \left( v_B^{pref} - \xi_{d(B)} \right),
    \end{equation}
    which using equation (\ref{equation:relativevelocitynew}) can be rewritten as
    \begin{equation} \label{equation:vabtrue}
        v_A - v_B =  - \xi_{d(A)} + \xi_{d(B)} + v_{AB}^{pref}.
    \end{equation}
    Since $\xi_{d(A)} \in \varepsilon_A$ and $\xi_{d(B)} \in \varepsilon_B$ from equation (\ref{equation:errorsinellipses}),
    \begin{equation} 
         - \xi_{d(A)} + \xi_{d(B)} \in -\varepsilon_A + \varepsilon_B,
    \end{equation}
    where 
    \begin{equation} \label{ellipsesymmetry}
        -\varepsilon_A = \varepsilon_A
    \end{equation}
    due to $\varepsilon_A$ being symmetric about the origin.
    Adding $v_{AB}^{pref}$ on both sides, we obtain
    \begin{equation} \label{equation:uncertainellipses}
         - \xi_{d(A)} + \xi_{d(B)} + v_{AB}^{pref} \in -\varepsilon_A + \varepsilon_B+v_{AB}^{pref}.
    \end{equation}
    Therefore, from equation (\ref{equation:vabtrue}) and (\ref{equation:uncertainellipses}),
    \begin{equation} \label{equation:proofwithminusea}
        v_A - v_B \in -\varepsilon_A + \varepsilon_B+v_{AB}^{pref}.
    \end{equation}
    Using (\ref{ellipsesymmetry}) allows us to rewrite equation (\ref{equation:proofwithminusea}) as
    \begin{equation} \label{equation:proofwithplusea}
        v_A - v_B \in \varepsilon_A + \varepsilon_B+v_{AB}^{pref}.
    \end{equation}
    Equations (\ref{minksumnottranslated}), (\ref{equation:minkowskisumellipses}), and (\ref{equation:proofwithplusea}) yield (\ref{equation:vaminusvbinepsilonab}).
\endproof

\begin{corollary} \label{corollary}
    Let $A$ and $B$ apply their control inputs $u_A$ and $u_B$ as defined in equation (\ref{equation:desiredcontrolinputwithwindrewritten}) with $i$ replaced by $A$ and $B$, respectively, based on given desired velocities $v_A^{pref}$ and $v_B^{pref}$. Let $v_A^{pref}$ and $v_B^{pref}$ be constant for a time horizon $\tau$. Assume that (\ref{equation:errorsinellipses}) is satisfied for both $\xi_{d(A)}$ and $\xi_{d(B)}$. Then $A$ and $B$ will not collide before $\tau$ units of time if
    \begin{equation} \label{equation(corollary):intersectionshapeandvo}
        \overline{\varepsilon_{AB}^{sum}} \cap \mathcal{VO}_{AB}^{\tau} = \emptyset.
    \end{equation}
\end{corollary}

\proof
    Using Theorem \ref{theorem:allvelocitiesineab} we know that 
    \begin{equation} \label{equation:truevelinminksum}
        v_{A} - v_{B} \in \overline{\varepsilon_{AB}^{sum}}.
    \end{equation}
    Moreover, Theorem \ref{theorem:collisioniffinVO} states that a collision between $A$ and $B$ happens if and only if
    \begin{equation}
        v_{A} - v_{B} \in \mathcal{VO}_{AB}^{\tau}.
    \end{equation}
    Thus, no collision occurs if and only if
    \begin{equation}
        v_{A} - v_{B} \notin \mathcal{VO}_{AB}^{\tau},
    \end{equation}
    which is the case when (\ref{equation(corollary):intersectionshapeandvo}) is satisfied.
\endproof

As a result of Corollary \ref{corollary}, we determine a vector $w$ that translates $\overline{\varepsilon_{AB}^{sum}}$ outside of $\mathcal{VO}_{AB}^{\tau}$. This is achieved by translating $\overline{\varepsilon_{AB}^{sum}}$ into a half-space $\pi$, which we define as
\begin{equation} \label{equation:halfplanetheorem6}
    \pi = \left\{ v \left| \left( v-p \right) \cdot n > 0 \right. \right\},
\end{equation}
where $p$ is computed as
\begin{equation} \label{eq:theorem6pdefinition_new}
    p = \underset{v \in \partial \mathcal{VO}_{AB}^{\tau}}{\text{argmin}} \norm{v-v_{AB}^{pref}}{},
\end{equation}
and $n$ is the outward normal vector of $\partial \mathcal{VO}_{AB}^{\tau}$ at $p$. Notice that
\begin{equation} \label{equation:PandVOdontintersecttheorem6}
    \pi \cap \mathcal{VO}_{AB}^{\tau} = \emptyset.
\end{equation}
We therefore solve
\begin{equation} \label{equation:shortestseperationvector}
    w= \left(
    \begin{split}
        &\underset{\bar{w}}{\text{argmin}}\norm{\bar{w}}{}\\
        &\textrm{s.t.} \left(\overline{\varepsilon_{AB}^{sum}}+\bar{w}\right) \in \pi\\
    \end{split}\right),
\end{equation}
or, equivalently,
\begin{equation} \label{walternativedefinition}
    w=\left[\left(p-x^*\right) \cdot n\right]  n,
\end{equation}
where
\begin{equation}
    x^*= \left(
    \begin{split}
        &\min n^Tx\\
        &\textrm{s.t.} x \in \overline{\varepsilon_{AB}^{sum}}
    \end{split}\right).
\end{equation}

\begin{figure}[t]
    \centering
    \includegraphics[scale=0.7]{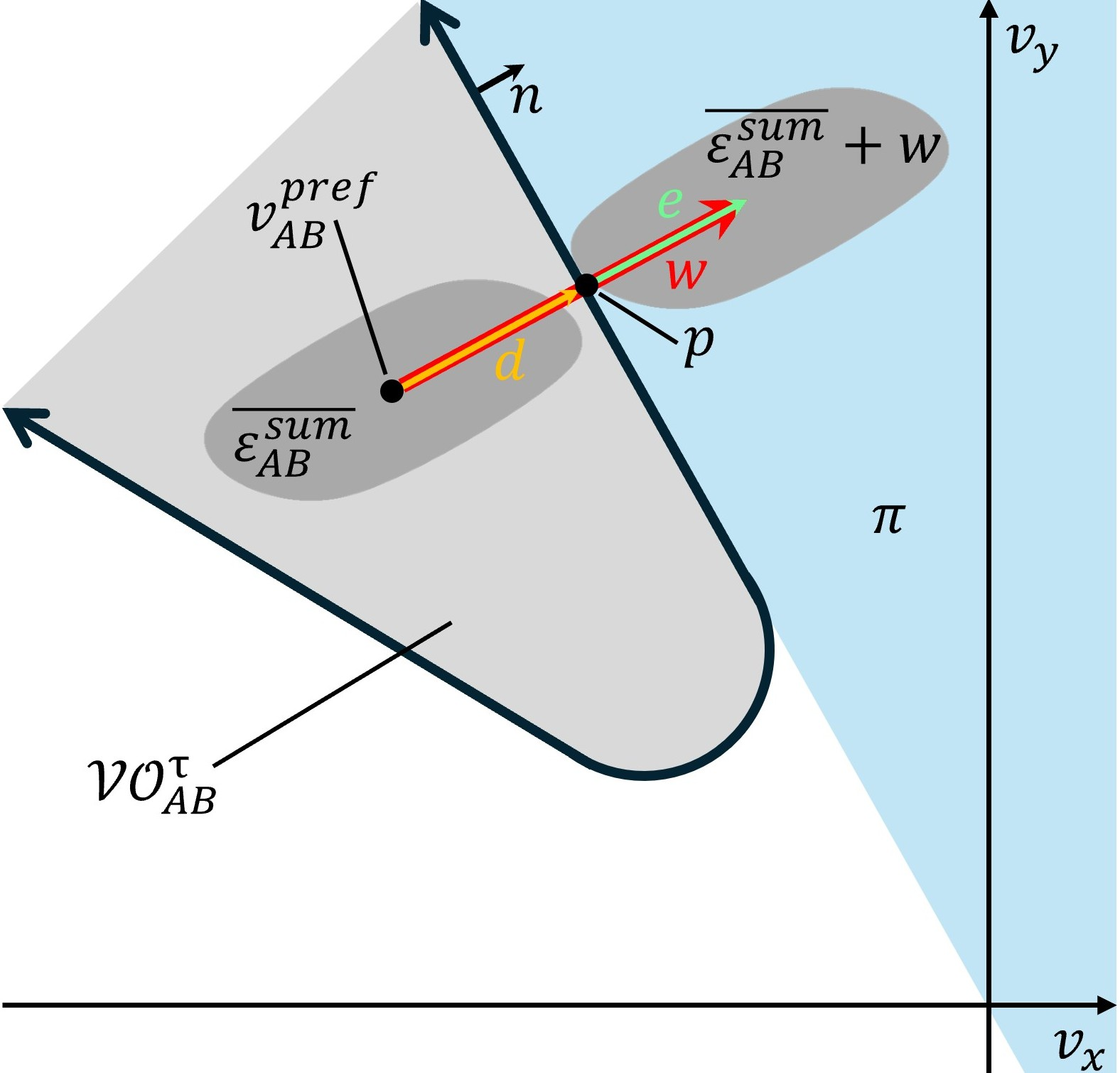}
    \caption{Visualization of $\pi$}
    \label{fig:theorem6}
\end{figure}

Note that following the \textit{Separating Hyperplane Theorem} (see reference \cite{boyd2004convex}, page 46) $\pi$ is the upper half-space of the hyperplane that separates $\overline{\varepsilon_{AB}^{sum}}+w$ from $\mathcal{VO}_{AB}^{\tau}$. A two-dimensional visualization of $w$ is given in figure \ref{fig:theorem6}. The translation vector $w$ is then used to define $ORCA_{AB}^{\tau}$ by equation (\ref{equation:halfplane}). After obtaining the half-spaces of permitted velocities with respect to all other agents, $A$ computes a set of collision-free velocities as in equation (\ref{equation:intersectionofhalfplanes}). The new velocity is obtained using equation (\ref{equation:ORCAnewvelocity}).

\begin{theorem} \label{theorem:multipleagentswithnoise}
    Let $A$ and $B$ apply their control inputs $u_A$ and $u_B$ as defined in equation (\ref{equation:desiredcontrolinputwithwindrewritten}) with $i$ replaced by $A$ and $B$, respectively, based on given velocities $v_A^{new} \in ORCA_{AB}^{\tau}$ and $v_B^{new} \in ORCA_{BA}^{\tau}$, where $ORCA_{AB}^{\tau}$ is computed as in equation (\ref{equation:halfplane}) and $ORCA_{BA}^{\tau}$ is computed as in equation (\ref{equation:orca_ba_halfplane}),
    with $w$ as defined in equation (\ref{equation:shortestseperationvector}) and where $n$ is the outward normal vector of $\partial \mathcal{VO}_{AB}^{\tau}$ at $p$ as defined in equation (\ref{eq:theorem6pdefinition_new}). Let $v_A^{new}$ and $v_B^{new}$ be constant for a time horizon $\tau$. Assume that the constraint (\ref{equation:errorsinellipses}) is satisfied for both $\xi_{d(A)}$ and $\xi_{d(B)}$. Then $A$ and $B$ do not collide before $\tau$, i.e.,
    \begin{equation}\label{notinset_newtheorem}
        v_A-v_B \notin \mathcal{VO}_{AB}^{\tau}.
    \end{equation}
\end{theorem}

\proof
    Using equation (\ref{truevelocitiesaandb}) with $v_A^{pref}$ replaced by $v_A^{new}$ and $v_B^{pref}$ replaced by $v_B^{new}$,
    \begin{equation} \label{equation:truevelwhenadoptingvAnew}
        v_A = v_A^{new} - \xi_{d(A)},
    \end{equation}
    \begin{equation} \label{equation:truevelwhenadoptingvBnew}
        v_B = v_B^{new} - \xi_{d(B)}.
    \end{equation}
    Choosing new velocities $v_A^{new} \in ORCA_{AB}^{\tau}$ (equation (\ref{equation:halfplane})) and $v_B^{new} \in ORCA_{BA}^{\tau}$ (equation (\ref{equation:orca_ba_halfplane})) must satisfy
    \begin{equation} \label{equation:theorem6:vAnewconstraint}
        \left( v_A^{new} - \left( v_A^{pref} + \frac{1}{2}w \right) \right) \cdot n > 0
    \end{equation}
    and
    \begin{equation} \label{equation:theorem6:vBnewconstraint}
        \left( v_B^{new} - \left( v_B^{pref} - \frac{1}{2}w \right) \right) \cdot \left(-n\right) > 0.
    \end{equation}
    Solving (\ref{equation:truevelwhenadoptingvAnew}) and (\ref{equation:truevelwhenadoptingvBnew}) for $v_A^{new}$ and $v_B^{new}$, respectively, and inserting them in (\ref{equation:theorem6:vAnewconstraint}) and (\ref{equation:theorem6:vBnewconstraint}) yields
    \begin{equation} \label{equation:theorem6:vAtrueconstraint}
        \left( v_A + \xi_{d(A)} - \left( v_A^{pref} + \frac{1}{2}w \right) \right) \cdot n > 0
    \end{equation}
    and
    \begin{equation} \label{equation:theorem6:vBtrueconstraint}
        \left( v_B + \xi_{d(B)} - \left( v_B^{pref} - \frac{1}{2}w \right) \right) \cdot \left(-n\right) > 0.
    \end{equation}
    Adding (\ref{equation:theorem6:vAtrueconstraint}) and (\ref{equation:theorem6:vBtrueconstraint}) gives
    \begin{equation} \label{eq:theorem6:ineqadded}
        \left( v_A - v_B + \xi_{d(A)} - \xi_{d(B)} - v_A^{pref} + v_B^{pref} - w \right) \cdot n > 0.
    \end{equation} 
    Using (\ref{equation:relativevelocitynew}) then yields
    \begin{equation} \label{eq:theorem6:ineqaddedsummarized}
        \left( v_A - v_B + \xi_{d(A)} - \xi_{d(B)} - v_{AB}^{pref} - w \right) \cdot n > 0.
    \end{equation}
    Define 
    \begin{equation} \label{ddefinition}
        d = p - v_{AB}^{pref}.
    \end{equation}
    Then $d$ and $n$ are collinear, i.e.,
    \begin{equation} \label{dandnparallel}
        d \parallel n \perp \partial \mathcal{VO}_{AB}^{\tau}.
    \end{equation}
    Let
    \begin{equation} \label{eq:theorem6edefinition}
        e = w-d,
    \end{equation}
    where $w$ is defined in (\ref{walternativedefinition}).
    Using equations (\ref{ddefinition}) and (\ref{eq:theorem6edefinition}), (\ref{eq:theorem6:ineqaddedsummarized}) can be rewritten as
    \begin{equation}
        \left( v_A - v_B + \xi_{d(A)} - \xi_{d(B)} - \left( p+e \right) \right) \cdot n > 0.
    \end{equation}
    Rearranging yields
    \begin{equation} \label{equation:inequalitytheorem6}
        \left( v_A - v_B - p \right) \cdot n > \left( e - \xi_{d(A)} + \xi_{d(B)} \right) \cdot n.
    \end{equation}
    From (\ref{dandnparallel}), (\ref{walternativedefinition}), and (\ref{eq:theorem6edefinition})
    \begin{equation}
        e = kn,~k>0.
    \end{equation}
    Since $\mathcal{VO}_{AB}^{\tau}$ is a closed set while $\pi$ is an open set,
    \begin{equation} \label{lengthofe}
        e \cdot n = \norm{e}{} > 0.
    \end{equation}
    Due to symmetry of $\varepsilon_{A}$ and $\varepsilon_{B}$,
    \begin{equation}
        \varepsilon_{AB}^{diff} = \varepsilon_A - \varepsilon_B = \varepsilon_A + (- \varepsilon_B) = \varepsilon_A +  \varepsilon_B = \varepsilon_{AB}^{sum},
    \end{equation}
    where $\varepsilon_{AB}^{sum}$ is defined in (\ref{minksumnottranslated}). Define an orthogonal coordinate system with origin at $v_{AB}^{pref}+w$ and with its primary axis along $n$. Consider the coordinate of $\xi_{d(A)}-\xi_{d(B)}$ along this axis, i.e., $(\xi_{d(A)}-\xi_{d(B)}) \cdot n$.  Then there are two cases,
    \begin{enumerate}
        \item $(\xi_{d(A)}-\xi_{d(B)}) \cdot n \leq 0$,
        \item $(\xi_{d(A)}-\xi_{d(B)}) \cdot n > 0$.
    \end{enumerate}
    Using (\ref{lengthofe}), for case 1 we can conclude that
    \begin{equation}
        \norm{e}{} = e \cdot n > (\xi_{d(A)}-\xi_{d(B)}) \cdot n.
    \end{equation}
    We will now prove by contradiction that this result still stands for case 2. To that aim, we first notice that, due to the symmetry of $\varepsilon_{AB}^{sum}$, for any $\xi_{d(A)}-\xi_{d(B)}$ in case 2 there is a corresponding point $\xi_{d(B)}-\xi_{d(A)}$ with negative coordinate along $n$, i.e.,
    \begin{equation}
        (\xi_{d(B)}-\xi_{d(A)}) \cdot n = -(\xi_{d(A)}-\xi_{d(B)}) \cdot n < 0.
    \end{equation}
    Assume that there are $\xi_{d(A)}^*$ and $\xi_{d(B)}^*$ in case 2 such that $(\xi_{d(A)}^*-\xi_{d(B)}^*) \cdot n \geq \norm{e}{}$. Then, this would imply that $\xi_{d(B)}^*-\xi_{d(A)}^* \in \mathcal{VO}_{AB}^{\tau}$ and therefore $\xi_{d(B)}^*-\xi_{d(B)}^* \notin \pi$, which violates the constraint in (\ref{equation:shortestseperationvector}). Note that the constraint in (\ref{equation:shortestseperationvector}) is on $\overline{\varepsilon_{AB}^{sum}}+w$, which is a shifted version of $\varepsilon_{AB}^{sum}$. From case 1 and case 2 we can conclude that
    \begin{equation} \label{equation:egreatersumdotproduct}
        e \cdot n > \left( \xi_{d(A)} - \xi_{d(B)} \right) \cdot n.
    \end{equation}
    Rearranging equation (\ref{equation:egreatersumdotproduct}) yields
    \begin{equation} \label{eq:noiseconditiontheorem6}
        \left( e - \xi_{d(A)} + \xi_{d(B)} \right) \cdot n > 0.
    \end{equation}
    Using equations (\ref{equation:inequalitytheorem6}) and (\ref{eq:noiseconditiontheorem6}) leads to
    \begin{equation} \label{equation_proof}
        \left( v_A - v_B - p \right) \cdot n > 0,
    \end{equation}
    From (\ref{equation_proof}) and (\ref{equation:halfplanetheorem6}),
    \begin{equation}
        v_A - v_B \in \pi.
    \end{equation}
    Then, using equation (\ref{equation:PandVOdontintersecttheorem6}) we conclude that (\ref{notinset_newtheorem}) is satisfied.
\endproof

\begin{corollary}
    Consider $m$ agents $A, B_1, ..., B_{m-1}$ operating within the same space. Let $A$ choose $v_A^{new} \in ORCA_{A}^{\tau}$, where
    \begin{equation}
        v_A = v_A^{new} - \xi_{d(A)}
    \end{equation}
    and $ORCA_{A}^{\tau}$ is defined in (\ref{equation:intersectionofhalfplanes}), with $ORCA_{AB_i}^{\tau}$ computed using $w$ as defined in (\ref{equation:shortestseperationvector}) and where $n$ is the outward normal vector of $\partial \mathcal{VO}_{AB}^{\tau}$ at $p$ as defined in (\ref{eq:theorem6pdefinition_new}). Then assuming that $v_A^{new}$ and $v_{B_i}^{new}$ are constant for a time horizon $\tau$ and that the constraint (\ref{equation:errorsinellipses}) is satisfied for both $\xi_{d(A)}$ and $\xi_{d(B_i)}$, $A$ will not collide with any other agent $B_i, i\in \left\{ 1,...,m-1 \right\}$ that chooses a new velocity $v_{B_i}^{new} \in ORCA_{B_iA}^{\tau}$, i.e.,
    \begin{equation} \label{corollary3:condition}
        v_A-v_{B_i} \notin \mathcal{VO}_{AB_i}^{\tau}, \forall i \in \left\{ 1,...,m-1 \right\},
    \end{equation}
    where, from equation (\ref{truevelocityredone}) with $i$ replaced by $B_i$, we get
    \begin{equation}
        v_{B_i} = v_{B_i}^{new} - \xi_{d(B_i)}.
    \end{equation}
\end{corollary}

\proof
    From the definition of $ORCA_{A}^{\tau}$ (equation (\ref{equation:intersectionofhalfplanes})), assuming that $v_A^{max}$ is finite,
    \begin{equation}
        ORCA_{A}^{\tau} \subset ORCA_{AB_i}^{\tau}, \forall i \in \left\{ 1,...,m-1 \right\}.
    \end{equation}
    Therefore, $v_A^{new} \in ORCA_A^{\tau}$ implies that $v_A^{new} \in ORCA_{AB_i}^{\tau}$. From Theorem \ref{theorem:multipleagentswithnoise}, we know that $v_A^{new} \in ORCA_{AB_i}^{\tau}$ and $v_{B_i}^{new} \in ORCA_{B_iA}^{\tau}$ leads to (\ref{corollary3:condition}). 
\endproof

\subsection{Application to Multi-Swarm Coverage Control} \label{subsection:algorithm}

\begin{algorithm}[t!]
    \caption{Collision Avoidance for Multi-Swarm Coverage Control with Disturbance Measurement Uncertainty}
    \begin{algorithmic}[1]
        \REQUIRE $X,Q,D$
	  \ENSURE $u_{i,j}^{new}$
            \STATE $X,D \leftarrow$ update($X,D$);
            \STATE $x_{i,j} \leftarrow X_j(i)$;
            \STATE $Edges_j,Vertices_j \leftarrow$ voronoi($X_j,Q$);
            \STATE $CM_{V_{i,j}} \leftarrow$ centerOfMass($Edges_j(i),Vertices_j(i)$);
            \STATE $v_{i,j}^{pref} \leftarrow$ desiredVelocity($CM_{V_{i,j}},x_{i,j},k_{i,j}$);
            \STATE $\hat{v}_{d(i,j)} \leftarrow$ estimateDisturbances(D);
            \FORALL{$X_l \subset X, \, l=\{1,...,N\} $} 
            \STATE $Edges_l,Vertices_l \leftarrow$ voronoi($X_l,Q$);
                \FORALL{$x_{k,l} \in X_l, \forall \, (k,l) \neq (i,j),k=\{1,...,m\}$}
                    \STATE $CM_{V_{k,l}} \leftarrow$ centerOfMass($Edges_l(k),$\\ \quad \quad \quad \quad \quad \quad \quad \quad \quad \quad \quad \quad $Vertices_l(k)$);
                    \STATE $\hat{v}_{k,l}^{pref} \leftarrow$ desiredVelocity($CM_{V_{k,l}},x_{k,l},\hat{k}_{k,l}$);
                    \STATE $\mathcal{VO}_{(i,j)(k,l)}^{\tau} \leftarrow$ velocityObstacle($x_{i,j},x_{k,l},$\\ \quad \quad \quad \quad \quad \quad \quad \quad \quad \quad \quad \quad \quad \quad \quad $r_{i,j},r_{k,l},\tau$);
                    \STATE $v_{(i,j)(k,l)}^{pref} \leftarrow v_{i,j}^{pref}-\hat{v}_{k,l}^{pref}$;
                    \STATE $\varepsilon_{i,j}, \varepsilon_{k,l} \leftarrow$ uncertaintyEllipsoids($\xi_{d(i,j)},\xi_{d(k,l)}$);
                    \STATE $\overline{\varepsilon_{(i,j)(k,l)}^{sum}} \leftarrow \varepsilon_{i,j} + \varepsilon_{k,l} + v_{(i,j)(k,l)}^{pref}$;
                    \STATE $ORCA_{(i,j)(k,l)}^{\tau} \leftarrow$ halfspace($\mathcal{VO}_{(i,j)(k,l)}^{\tau},$\\ \quad \quad \quad \quad \quad \quad \quad \quad \quad \quad \quad \quad \quad $\overline{\varepsilon_{(i,j)(k,l)}^{sum}},v_{(i,j)(k,l)}^{pref}$);
                    \STATE $ORCA_{(i,j)(l)}^{\tau} \leftarrow \bigcap_{k=1}^{m} ORCA_{(i,j)(k,l)}^{\tau}$;
                \ENDFOR
                \STATE $ORCA_{i,j}^{\tau} \leftarrow D(0,v_{i,j}^{max}) \cap  \bigcap_{l=1}^{N-1} ORCA_{(i,j)(l)}^{\tau}$;
            \ENDFOR
            \STATE $v_{i,j}^{new} \leftarrow$ newVelocity($ORCA_{i,j}^{\tau}$);
            \STATE $u_{i,j}^{new} \leftarrow v_{i,j}^{new}-\hat{v}_{d(i,j)}$;
        \RETURN{} $u_{i,j}^{new}$
     \end{algorithmic}
     \label{algorithm:u_new}
\end{algorithm}

The proposed algorithm to integrate the collision avoidance method presented in subsection \ref{subsection:math} and coverage control with multiple swarms, is shown in table \ref{algorithm:u_new}. We make the following assumptions:

\begin{figure*} [t]
    \centering
    \subfloat[]{\includegraphics[width=0.5\textwidth]{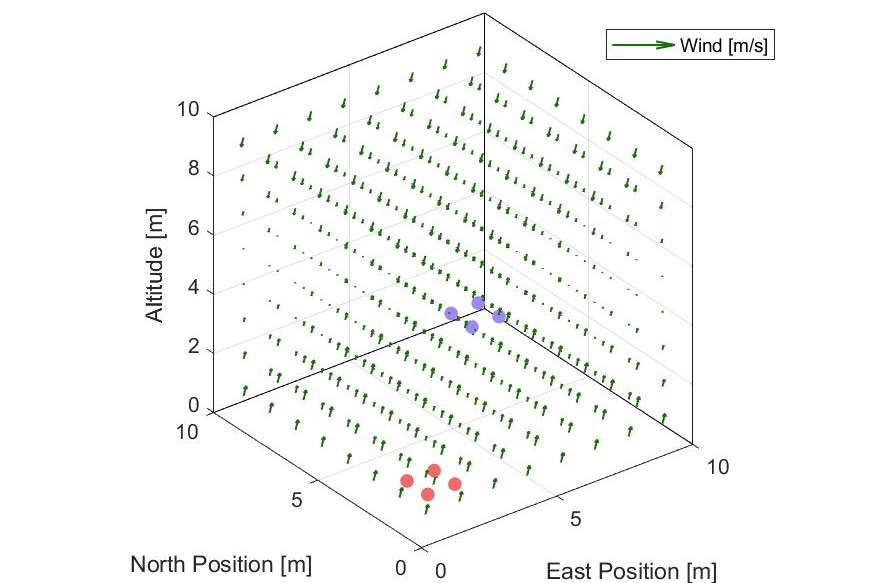}}
    \subfloat[]{\includegraphics[width=0.5\textwidth]{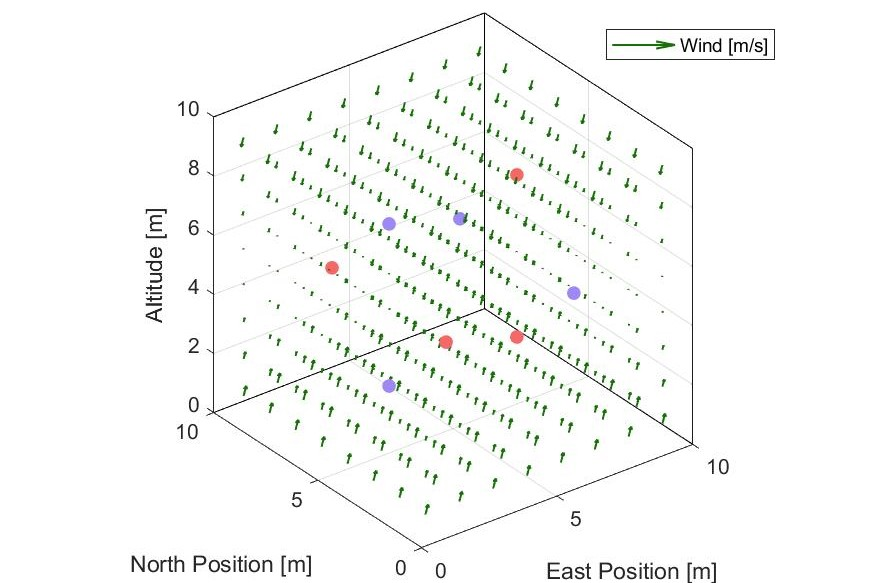}}

    \caption{Positions of the agents of swarm 1 (S1) and swarm 2 (S2) (a) at the beginning and (b) after 5 seconds.}
    \label{fig:simulation_results}
\end{figure*}

\begin{enumerate}
    \item All swarms use the same algorithm to cover the region and avoid other vehicles.
    \item The computations are executed synchronously for all agents.
    \item Group affiliation, as well as the radius of the sphere that defines a safety zone around an agent, are known for each agent.
    \item Knowledge regarding the absolute positions of all other agents within the same swarm can be obtained through communication between vehicles.
    \item The agents have sufficient sensing capabilities and communication within each swarm to determine the absolute positions of all agents from other swarms.
    \item The input needed to counteract disturbances can always be achieved.
\end{enumerate}

\begin{remark}
    Note that assumptions 1-4 are commonly used in multi-agent systems \cite{mao2022assumption1,assumption2,assumption3,assumption4}. For assumption 5, if there are at least 4 agents within one swarm and their relative positions are linearly independent, range measurements can be used to determine the positions of the agents from other swarms. Note that if assumption 6 is not satisfied, it is pointless to design a control input that counteracts disturbances.
\end{remark}

The algorithm is run for each agent. The agent for which the algorithm is run is denoted as agent $i$ belonging to swarm $j$. This is indicated by the subscript $i,j$. Note that $X$ contains the positions of all agents, while $D$ contains the disturbance measurements.

\section{Simulations} \label{section:simulations}

To validate the results, this section shows simulations of the kinematics of two swarms of fully actuated UAVs that independently aim to cover a three-dimensional region subject to a wind field. The region to cover is defined as a cube with an edge length of $10$ meters (m) and has a uniform density distribution. We define a coordinate system with the origin at one vertex of the cube and the axes pointing in the East ($x$), North ($y$), and Up ($z$) directions, parallel to the faces of the cube. Both swarms consist of four agents initially located at
\begin{enumerate}
     \item $[(1,1,1),(2,2,1),(2,1,1),(1,2,1)]$m,
     \item $[(9,9,1);(8,8,1);(8,9,1);(9,8,1)]$m,
\end{enumerate}
as illustrated in figure \ref{fig:simulation_results}a. The maximum speed is set to $5$ meters per second (m/s) for all agents. We define the wind field as $F(x,y,z) = \left( 4-z,4-z,0\right)$m/s, which corresponds to shear wind. For the wind measurement, the ellipsoid $\varepsilon_i$ that bounds $\xi_d(i,t)=\delta b_d{(i)}+e_{d(i,t)}$ is defined by its semi-major axis $a=0.15v_{d(i,t)}$ which is aligned with the direction of the wind and the two perpendicular semi-minor axes $b$ and $c$ with norms $\norm{b}{}=\norm{c}{}=\norm{0.005v_{d(i,t)}}{}$. The time horizon is chosen to be $\tau = 1$ second (s), and the time step is $T=0.01$s. Figure \ref{fig:simulation_results}b shows the locations of the agents after $5$s, which is after the agents reach their final destinations. It can be seen that the agents from each swarm spread out during the simulation to cover the region. In particular, after $5$s, the volume of the Voronoi cells of all agents is within $0.001 \%$ of $250 m^3$. Since $250 m^3$ is the total volume of the region to be covered divided by $4$, which is the number of agents per swarm, this indicates that locally optimal coverage is achieved by both swarms. Furthermore, throughout the entire simulation, no collision occurred. 

\begin{figure} [t]
    \subfloat[]{\includegraphics[width=0.5\textwidth]{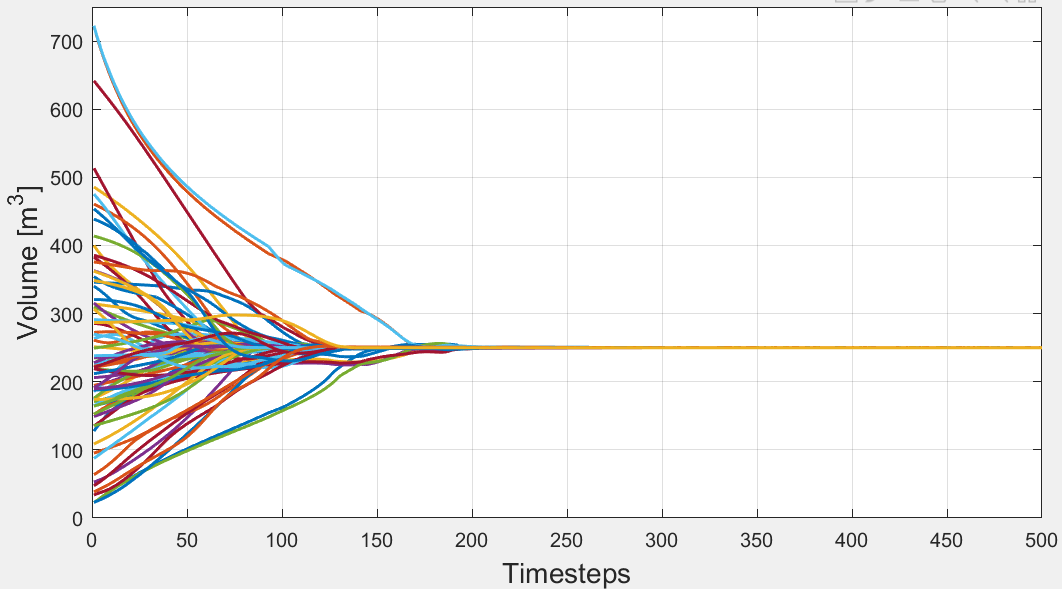}}
    \newline
    \subfloat[]{\includegraphics[width=0.5\textwidth]{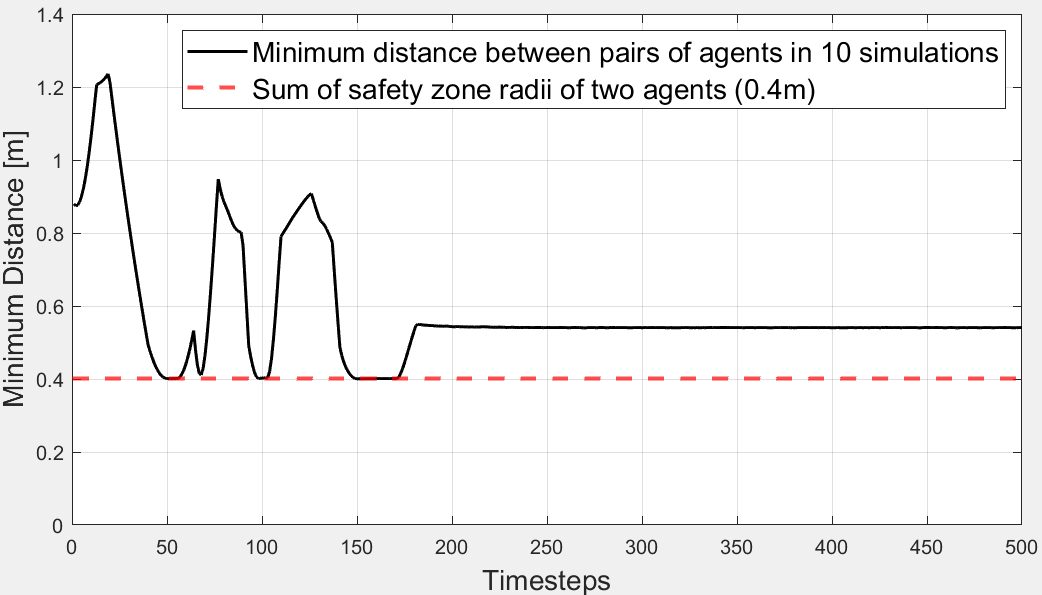}}

    \caption{Volume of Voronoi cells of each agent (a) and distance between each pair of agents from different swarms (b)}
    \label{fig:simulation_volume_distance}
\end{figure}

Figure \ref{fig:simulation_volume_distance} summarizes the results of 10 simulation runs for the same parameters, but with the agents starting from random initial positions. In figure \ref{fig:simulation_volume_distance}a, the volume of the Voronoi cell of each agent is plotted over time. It can be seen that at the beginning, some agents cover a large volume, while others have less volume to cover. However, at the end of all simulations, the volumes of the Voronoi cells of all agents converge to $250 m^3$, which suggests locally optimal coverage in all simulations. Moreover, figure \ref{fig:simulation_volume_distance}b visualizes the minimum of the distance between each pair of agents belonging to different swarms at each timestep during all 10 simulations. Since the minimum distance never falls below the sum of the radii of the safety zones of two agents ($0.4$m), we can conclude that no collision occurs. These results demonstrate the effectiveness of the proposed methodology and algorithm, as well as their applicability to real-world scenarios. 

\section{Conclusions} \label{section:conclusions}
This paper introduced a collision avoidance methodology for multi-agent systems. The method specifically accounts for bounded disturbances and uncertainty in their measurements, which makes it applicable to real-world environments. Since guarantees for collision-free motion in coverage control with only one swarm do not hold for the multi-swarm case, an algorithm was presented that introduces the use of the proposed methodology for collision-free motion in multi-swarm coverage control. Collision avoidance is formally guaranteed. Simulations showed that using the new algorithm, both collision-free motion and locally optimal coverage can be achieved by multiple swarms operating within a three-dimensional region.

\bibliographystyle{IEEEtran}
\bibliography{IEEEabrv,References}

\end{document}